\documentclass[twoside,twocolumn,9pt]{article}
\usepackage{extsizes}
\usepackage[super,sort&compress,comma]{natbib} 
\usepackage[version=3]{mhchem}
\usepackage[left=1.5cm, right=1.5cm, top=1.785cm, bottom=2.0cm]{geometry}
\usepackage{balance}
\usepackage{mathptmx}
\usepackage{sectsty}
\usepackage{graphicx} 
\usepackage{lastpage}
\usepackage[format=plain,justification=justified,singlelinecheck=false,font={stretch=1.125,small,sf},labelfont=bf,labelsep=space]{caption}
\usepackage{float}
\usepackage{fancyhdr}
\usepackage{fnpos}
\usepackage[english]{babel}
\addto{\captionsenglish}{%
  
}
\usepackage{array}
\usepackage{droidsans}
\usepackage{charter}
\usepackage[T1]{fontenc}
\usepackage[usenames,dvipsnames]{xcolor}
\usepackage{setspace}
\usepackage[compact]{titlesec}
\usepackage{hyperref}
\usepackage{epstopdf}%This line makes .eps figures into .pdf - please comment out if not required.

\usepackage{amsmath,amssymb}
\usepackage{multirow}

\definecolor{cream}{RGB}{222,217,201}

\begin{document}

\pagestyle{fancy}
\thispagestyle{plain}
\fancypagestyle{plain}{
%%%HEADER%%%
\renewcommand{\headrulewidth}{0pt}
}
%%%END OF HEADER%%%

%%%PAGE SETUP - Please do not change any commands within this section%%%
\makeFNbottom
\makeatletter
\renewcommand\LARGE{\@setfontsize\LARGE{15pt}{17}}
\renewcommand\Large{\@setfontsize\Large{12pt}{14}}
\renewcommand\large{\@setfontsize\large{10pt}{12}}
\renewcommand\footnotesize{\@setfontsize\footnotesize{7pt}{10}}
\makeatother

\renewcommand{\thefootnote}{\fnsymbol{footnote}}
\renewcommand\footnoterule{\vspace*{1pt}% 
\color{cream}\hrule width 3.5in height 0.4pt \color{black}\vspace*{5pt}} 
\setcounter{secnumdepth}{5}

\makeatletter 
\renewcommand\@biblabel[1]{#1}            
\renewcommand\@makefntext[1]% 
{\noindent\makebox[0pt][r]{\@thefnmark\,}#1}
\makeatother 
\renewcommand{\figurename}{\small{Fig.}~}
\sectionfont{\sffamily\Large}
\subsectionfont{\normalsize}
\subsubsectionfont{\bf}
\setstretch{1.125} %In particular, please do not alter this line.
\setlength{\skip\footins}{0.8cm}
\setlength{\footnotesep}{0.25cm}
\setlength{\jot}{10pt}
\titlespacing*{\section}{0pt}{4pt}{4pt}
\titlespacing*{\subsection}{0pt}{15pt}{1pt}
%%%END OF PAGE SETUP%%%

%%%FOOTER%%%
\fancyfoot{}
\fancyfoot[LO,RE]{\vspace{-7.1pt}\includegraphics[height=9pt]{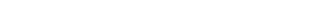}}
\fancyfoot[CO]{\vspace{-7.1pt}\hspace{13.2cm}\includegraphics{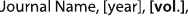}}
\fancyfoot[CE]{\vspace{-7.2pt}\hspace{-14.2cm}\includegraphics{head_foot/RF}}
\fancyfoot[RO]{\footnotesize{\sffamily{1--\pageref{LastPage} ~\textbar  \hspace{2pt}\thepage}}}
\fancyfoot[LE]{\footnotesize{\sffamily{\thepage~\textbar\hspace{3.45cm} 1--\pageref{LastPage}}}}
\fancyhead{}
\renewcommand{\headrulewidth}{0pt} 
\renewcommand{\footrulewidth}{0pt}
\setlength{\arrayrulewidth}{1pt}
\setlength{\columnsep}{6.5mm}
\setlength\bibsep{1pt}
%%%END OF FOOTER%%%

%%%FIGURE SETUP - please do not change any commands within this section%%%
\makeatletter 
\newlength{\figrulesep} 
\setlength{\figrulesep}{0.5\textfloatsep} 

\newcommand{\topfigrule}{\vspace*{-1pt}% 
\noindent{\color{cream}\rule[-\figrulesep]{\columnwidth}{1.5pt}} }

\newcommand{\botfigrule}{\vspace*{-2pt}% 
\noindent{\color{cream}\rule[\figrulesep]{\columnwidth}{1.5pt}} }

\newcommand{\dblfigrule}{\vspace*{-1pt}% 
\noindent{\color{cream}\rule[-\figrulesep]{\textwidth}{1.5pt}} }

\makeatother
%%%END OF FIGURE SETUP%%%

%%%TITLE, AUTHORS AND ABSTRACT%%%
\twocolumn[
  \begin{@twocolumnfalse}

\vspace{1em}
\sffamily
\begin{tabular}{m{4.5cm} p{13.5cm} }

\includegraphics{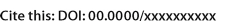} & \noindent\LARGE{\textbf{Stochastic migrations of Marangoni surfers between two lobes of a dumbbell-shaped confinement$^\dag$}} \\%Article title goes here instead of the text "This is the title"
\vspace{0.3cm} & \vspace{0.3cm} \\

& \noindent\large{Alakesh Upadhyaya\textit{$^{a}$} and V.S. Akella$^{\ast}$\textit{$^{a}$}} \\%Author names go here instead of "Full name", etc.

\includegraphics{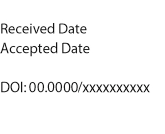} & \noindent\normalsize{We report an experimental investigation on the stochastic migration dynamics of Marangoni surfers (camphor-infused paper disks) between the two lobes of a dumbbell-shaped chamber. We characterize the migration dynamics using survival analysis for a given configuration with a unique disk count in each lobe. Qualitatively, all the configurations exhibit stretched exponential decay with time, ascribed to a disk's ``aging''. A given configuration's stability decreases with increasing pairwise interactions. The most stable configuration is \emph{always} the one with equal partitioning of disks between the lobes, i.e., exactly one-half of disks in each lobe for even-numbered systems but with one extra disk in either of the lobes for odd-numbered systems. Furthermore, we model a camphor disk as a Chiral Active Particle (CAP), as initially proposed by Cruz et al.~\cite{Cruz2024-SoftMatter}, and show that decreasing motility (or aging) indeed causes the stretched exponential behavior. } \\%The abstract goes here instead of the text "The abstract should be..."

\end{tabular}

 \end{@twocolumnfalse} \vspace{0.6cm}

]
%%%END OF TITLE, AUTHORS AND ABSTRACT%%%

%%%FONT SETUP - please do not change any commands within this section
\renewcommand*\rmdefault{bch}\normalfont\upshape
\rmfamily
\section*{}
\vspace{-1cm}

%%%FOOTNOTES%%%

\footnotetext{\textit{$^{a}$~Department of Physics, Indian Institute of Technology Jammu, NH-44, Jagti Village, Jammu, J\& K, India. Tel: +91 191 257 1069; $^{\ast}$~\textsf{sathish.akella@iitjammu.ac.in}}}

%Please use \dag to cite the ESI in the main text of the article.
%If you article does not have ESI please remove the the \dag symbol from the title and the footnotetext below.

%additional addresses can be cited as above using the lower-case letters, c, d, e... If all authors are from the same address, no letter is required

% \footnotetext{\ddag~Additional footnotes to the title and authors can be included \textit{e.g.}\ `Present address:' or `These authors contributed equally to this work' as above using the symbols: \ddag, \textsection, and \P. Please place the appropriate symbol next to the author's name and include a \texttt{\textbackslash footnotetext} entry in the the correct place in the list.}

%%%END OF FOOTNOTES%%%

%%%MAIN TEXT%%%%
\section{Introduction}
Unlike conventional matter, which primarily adheres to principles of equilibrium statistical physics, active matter comprises agents/entities that derive energy from the surroundings to remain ``active'' and thus is out of equilibrium. Common examples of active matter span various lengths and time scales including macroscopic living matter such as flocks of birds~\cite{Cavagna2014-ARCMP}, biological microswimmers~\cite{Tung2017-SR}, synthetic colloidal motors~\cite{Zhihua2018-CPCIS}, vibrated granular media~\cite{Kumar2014-NC}, and chemotactic surfers like Janus particles~\cite{Vutukuri2020-NC}.

A common characteristic among active matter systems is the complex interplay between the agents' activity and the surroundings that manifest into collective patterns at the macroscopic level, which are further affected by the intrinsic nature of agents, their number density, and the surrounding environment~\cite{Marcheti2013-RMP, Bar2020-ARCMP}. Further, one significant factor contributing to the emergent dynamics in active systems is \emph{confinement}. Investigating the dynamics of active matter within various confined geometries has become a significant area of interest over the past few decades~\cite{Tailleur2008-PRL, Takatori2016-NC, Hagan2014-SM, Yang2014-SM, Sokolov2007-PRL, Hernandez2005-PRL, Snigdhathakur2012-PRE, Lushi2014-PNAS}. Recent simulations and experiments \cite{Hagan2014-SM, Wioldand2013-PRL, Lushi2014-PNAS} have demonstrated that appropriately designed boundaries can stabilize and control order in active bacterial suspensions. Similarly, Active Brownian Particles (ABP) simulated under annular confinement~\cite{Caprini2021-JCP} have revealed collective rotations despite the absence of explicit alignment interactions. Furthermore, recent simulations~\cite{Paoluzzi2020-PRE} have demonstrated that a mixture of active particles with different motilities can be effectively segregated by confining them to a chamber featuring a narrow escape aperture, as evidenced by the respective escape times. In biological systems, often, such apertures correspond to narrow paths or channels through which an agent can migrate, a phenomenon ubiquitous in nature. Investigating such migration dynamics can offer insights ranging from understanding cell dynamics at the molecular level to the movements of viruses and bacteria under such constraints. For instance, experimental studies~\cite{Ram2012-PNAS} have revealed that epithelial cells migrate when confined within rectangular strips of varying widths, with migration speed increasing as the strip width decreases, highlighting the significant role of constraints in cell motility. In addition, Bruckner et al.~\cite{Bruckner2019-NP} observed that a single cell exhibited stochastic migration when confined within a two-state micro-pattern comprising two adhesive sites connected by a thin constriction. However, when the constriction was removed, the nature of the migration dynamics of the cells underwent few changes. This study also showed that a cancerous cell exhibits a limit-cycle when confined to such an environment, while a non-cancerous cell exhibits bistable dynamics under the same conditions. Similarly, a simulation of an active bacterial suspension~\cite{Paoluzzi2015-PRL} under a dumbbell-shaped confinement displayed a regular density oscillation of active particles between the chambers. This simulation was later complemented by an experiment~\cite{Li2022-SM} where self-propelled granular particles demonstrated similar collective oscillatory behavior when confined to such chambers. Further, Biswas et al.~\cite{Biswas-SM-2023} studied the escape dynamics of an undulating worm from a 2D circular domain with an escape aperture and highlighted the boundary-tracing strategy of the worm to escape the domain.

Regardless of the insights gained from these studies, there remains a notable knowledge gap concerning the dynamics of multi-particle systems (both interacting and non-interacting) subjected to different types of confinement. To this end, we investigate the migration dynamics of Marangoni surfers between the two lobes of a dumbbell-shaped confinement. We employ camphor-infused paper disks as model Marangoni surfers, because of\textemdash the extensive literature~\cite{Akella2018-JPLA, Ishant2023-SM, NAGAYAMA2004-PhysicaD, Nakata2004-PCCP, Nakata2004-PRE, Nakata2022-CSA, NAKATA2022-Colloids, Nakats2021-JPCB, Nakata2020-JCPB} on camphor-infused disks under different experimental scenarios and the ease of preparation. We characterize the migration dynamics using survival analysis of a particular configuration with a unique disk count in each lobe. Further, we qualitatively capture some of the observations by simulating a camphor boat as a Chiral Active Particle (CAP), as proposed initially by Cruz et al.~\cite{Cruz2024-SoftMatter} 

The structure of the paper is as follows. In Section~\ref{Methods}, we elaborate on the experimental methods employed in this study. Section~\ref{Results} presents the findings related to both single-disk dynamics and multi-disk dynamics. Finally, in Section~\ref{Conclusion}, we present concluding remarks, highlight some questions, and suggest potential avenues for further research on similar problems.

\section{Experimental setup and methods}
\label{Methods}

\begin{figure}[t]
\includegraphics[width=\linewidth]{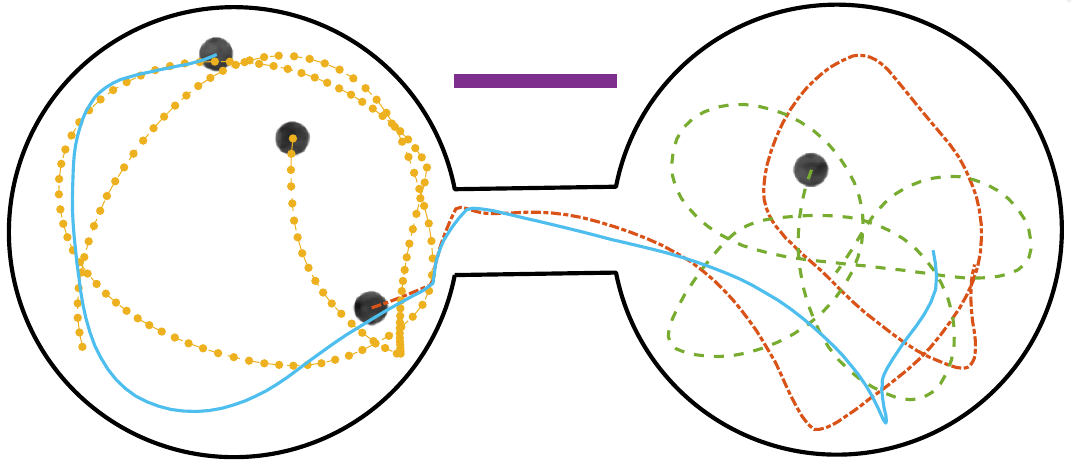}
\caption{Top view (contrast adjusted and dumbbell boundary highlighted for clarity) of the petri-dish with camphor disks ($\varnothing = 6\ \mathrm{mm}$) on water surface. The disks are bounded by a two-dimensional dumbbell-shaped confinement which freely-floats on water surface. The petri-dish is illuminated from the bottom side with a uniformly lit LED tablet. Scale Bar = $2.5$ \emph{cm} }
  \label{fig:figure1}
\end{figure}

We fabricate the camphor disks (generally called \emph{Marangoni surfers}) by following the procedure outlined in Ishant et al. ~\cite{Ishant2023-SM}. The procedure involves first punching out paper disks (diameter $\varnothing = 6\ \mathrm{mm}$) from a bond paper using a hole puncher and then soaking these in $1\ \mathrm{g/ml}$ camphor+ethanol solution overnight. Before starting an experiment, the soaked disks are dried in ambient air (at a room temperature of 26 \textcelsius) for $15-20\ \mathrm{min}$ to facilitate the evaporation of ethanol and deposition of camphor into the paper's fiber matrix. Henceforth the camphor disks are assumed to be \emph{identical} and \emph{indistinguishable}. We used research-grade camphor and ethanol for the procedure. The experimental setup consists of a circular petri dish ($\varnothing = 18\ \mathrm{cm}$), filled with $450\ \mathrm{ml}$ of de-ionized water (resistivity $\rho = 18.2\ \mathrm{M\Omega\ cm}$), placed on a uniformly lit square shaped LED tablet. In addition, we freely floated a circular ($\varnothing \lessapprox 18\ \mathrm{cm}$ and thickness $2\ \mathrm{mm}$) acrylic sheet with a dumbbell-shaped opening on the water surface. For stability, we firmly fastened the acrylic sheet in position. The dumbbell-shaped slot is cut out precisely into shape using a $\mathrm{CO}_{\mathrm{2}}$-laser cutter. The two lobes of the dumbbell opening are circular with $5\ \mathrm{cm}$ in diameter and are interconnected by a rectangular channel of length $2.5\ \mathrm{cm}$ and width $1.2\ \mathrm{cm}$. The dumbbell chamber dimensions are carefully chosen to allow interactions among disks in a lobe and smooth passage between the lobes. At the beginning of an experiment, camphor disks were gently introduced on the water surface in either lobe and their motion was recorded using an overhead DSLR camera (Nikon D5600) at 60 fps (1080p) for $20\ \mathrm{min}$. Figure~\ref{fig:figure1} shows a snapshot of four camphor disks on water surface bounded by a dumbbell-shaped confinement. The recorded motion of the camphor disks was tracked and analyzed using a particle tracking algorithm in MATLAB\textsuperscript{\textregistered} developed by Daniel Blair and Eric Dufresne~\cite{Blair}, which was originally based on a particle tracking algorithm written by John C. Crocker and David G. Grier~\cite{CROCKER1996-JCI}.

\begingroup
\setlength{\tabcolsep}{8pt} % Default value: 6pt
\renewcommand{\arraystretch}{1.2} % Default value: 1
\begin{table}[h!]
\begin{center}
\begin{tabular}{|c|c|c|c|}
\hline
N & Configuration & Label \\
\hline\hline
1 & $\langle$\textbullet \textbar$\rangle$ & $S_{1}^{1}$ \\
\cline{2-2} & $\langle$\textbar \textbullet$\rangle$ & \\ 
\hline\hline
2 & $\langle$\textbullet \textbullet \textbar$\rangle$ & $S_{2}^{2}$ \\
\cline{2-2} &  $\langle$\textbar \textbullet \textbullet$\rangle$ &  \\ 
\cline{2-3} &  $\langle$\textbullet \textbar \textbullet$\rangle$ & $S_{1}^{2}$ \\
\hline\hline
3 & $\langle$\textbullet \textbullet \textbullet \textbar$\rangle$& $S_{3}^{3}$ \\
\cline{2-2} & $\langle$\textbar \textbullet \textbullet \textbullet$\rangle$  & \\ 
\cline{2-3} & $\langle$\textbullet \textbullet \textbar \textbullet$\rangle$ & $S_{2}^{3}$ \\
\cline{2-2} & $\langle$\textbullet \textbar \textbullet \textbullet$\rangle$  & \\
\hline \hline
4 & $\langle$\textbullet \textbullet \textbullet \textbullet \textbar$\rangle$ & $S_{4}^{4}$ \\
\cline{2-2} & $\langle$\textbar \textbullet \textbullet \textbullet \textbullet$\rangle$  & \\ 
\cline{2-3} & $\langle$\textbullet \textbullet \textbullet \textbar \textbullet$\rangle$ & $S_{3}^{4}$ \\
\cline{2-2} & $\langle$\textbullet \textbar \textbullet \textbullet \textbullet$\rangle$ & \\
\cline{2-3} & $\langle$\textbullet \textbullet \textbar \textbullet \textbullet$\rangle$ & $S_{2}^{4}$ \\
\hline
\end{tabular}
\caption{List of possible configurations and their corresponding complementary configurations for \emph{identical} and \emph{indistinguishable} disks undergoing migrations between the lobes of the dumbbell-shaped confinement. ``\textbullet'' symbol represents a disk and ``\textbar'' symbol acts as a separator between the left and right lobes. For example, $\langle$\textbullet \textbar \textbullet \textbullet$\rangle$ represents a three-disk system with one disk in the left lobe and two disks in the right lobe, which also has a complementary configuration $\langle$\textbullet \textbullet \textbar \textbullet$\rangle$ with two disks in the left and one disk in right lobes. Both of these configurations are grouped together and corresponding survival probability is labeled $S_{2}^{3}$.}
\label{tab:table1}
\end{center}
\end{table}
\endgroup

\begin{figure}[h!]
\includegraphics[width=\linewidth]{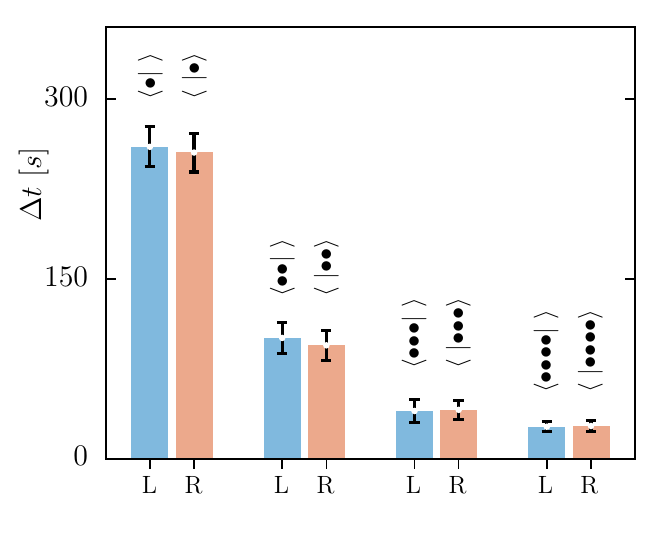}
\caption{The average time (measured from 14 trials) spent in a given configuration over an experimental duration of approximately 10 minutes\textemdash clearly showed no preference between the lobes in both single- and multi-disk experiments. The labels `L' for left and `R' for right are arbitrary.}
  \label{fig:figure2}
\end{figure}

When a camphor disk is placed on the water surface, it is spontaneously set into motion by the surface tension gradients, which in turn are caused by the release of camphor onto the water surface. Moreover, it stochastically shuttles back and forth between the lobes (see supplementary movie \emph{SM-1P.mp4}) after dwelling in a particular lobe for a duration and is aptly referred to as the \emph{dwell time}. For simplicity, we consider the mid-point between the lobes as the reference to determine the lobe occupancy, i.e., if the disk is to the left of the mid-point, then it is in the left lobe; otherwise in the right. We verified that considering lobe openings (separately for both the lobes) as references to calculate the dwell times (and the corresponding distribution) does not significantly alter the results. Now, we calculate the dwell time distribution for a \emph{configuration} and not for an individual disk\textemdash wherein a configuration describes a unique disk count in each lobe. Note that one can rather calculate individual dwell time distributions for each disk, however, when interactions exist among the disks, dwell time distribution for a configuration is a more relevant measure. See table~\ref{tab:table1} for a list of possible configurations in different scenarios. Figure~\ref{fig:figure2} is a bar plot of the average time spent in a given configuration over an experimental duration (approximately 10 minutes). It is evident from the figure that mutually complementary configurations subsist for approximately the same duration. Further, we performed the two-sample Kolmogorov-Smirnov test~\cite{Massey-JASA-1951} to confirm that the dwell times calculated separately for each lobe are, in fact, generated from the same underlying probability distribution. Therefore, we assume that there is \emph{no} preference between the two lobes of the dumbbell-shaped chamber and calculated the distribution of the dwell times by combining the statistics from both the lobes in single- as well as multi-disk experiments.

In the case of single disk experiments, a disk being in, say, the left lobe $\langle$\textbullet \textbar$\rangle$ is a \emph{complementary} state of being in the right $\langle$\textbar \textbullet$\rangle$, and the dwell time distributions for the left and right lobes are identical because there is \emph{no} particular preference between the lobes. Similarly, in a three-disk system, a configuration with two disks in the left lobe and one disk in the right lobe $\langle$\textbullet \textbullet \textbar \textbullet$\rangle$ has a complementary configuration $\langle$\textbullet \textbar \textbullet\textbullet$\rangle$ and the corresponding dwell time distributions are identical.

\paragraph*{}
\begin{equation} \label{eq:survive}
S_{\#}^{N}(t) = 1- \int_{0}^{t} P_{\#}^{N}(\tau) \mathrm{d}\tau 
\end{equation}

Now to analyze the migration dynamics, we calculate the survival probability $S_{\#}^{N}(t)$ ($\# = 1,2,\dots N$) for a configuration in a $N$-disk experiment from the corresponding dwell time distribution $P_{\#}^{N}$ using eqn.~\ref{eq:survive}. In general, $S_{\#}^{N}(t)$ is also known as complementary cumulative distribution function. For example, $S_{2}^{3}(t)$ represents the survival probability for two disks to simultaneously dwell in a lobe, i.e., $\langle$\textbullet \textbullet \textbar \textbullet$\rangle$ or $\langle$\textbullet \textbar \textbullet\textbullet$\rangle$ in a three-disk experiment. Further, for a homogeneous Poisson process $S_{\#}^{N}(t)$ is a decaying exponential. 

\section{Results and discussion}
\label{Results}
Camphor is a waxy, aromatic organic compound. When added to water, it reduces the surface tension of water; however, it is not regarded as a surfactant as it lacks the head and tail groups typical of a surfactant. The surface tension of pure water is $\sim 72\times10^{-3}\ \mathrm{N\cdot m^{-1}}$, where as water saturated with camphor is $\sim 60\times10^{-3}\ \mathrm{N\cdot m^{-1}}$ at 25\textcelsius~\cite{Akella2018-JPLA}. When a camphor disk (or any irregular-shaped fragment) is placed on a water surface, camphor readily diffuses onto the surface, creating a camphor-adsorbed region around the disk with radially varying camphor concentration determined by the balance between the influx of camphor onto the surface and outflux due to both dissolution into bulk water and sublimation in surrounding air~\cite{Akella2018-JPLA}. In the steady state, the established camphor concentration field (and hence the surface tension field) around the disk is such that the net thrust on the disk is zero. However, ambient noise, such as temperature fluctuations, non-uniform dissolution of camphor, etc, quickly destabilizes the net-zero-thrust state, and the disk is set into motion towards the high surface tension (or low camphor concentration) region. Additionally, the thrust is not always necessarily directed along the center of mass of the disk, thereby generating a torque on the disk. Thus, the disk's overall motion trajectory generally comprises curved paths with rectilinear sections (see figure~\ref{fig:figure1}). Usually, the activity (i.e., the self-propulsion strength) of a camphor disk gradually decreases over time due to the continuous adsorption (on water surface) and/or dissolution (in bulk water) of camphor, thereby gradually decreasing the overall surface tension. As a result the disk exhibits various modes of motion~\cite{Akella2018-JPLA, tiwari2021-SM}. Further, the thrust becomes zero when the camphor concentration on/in water reaches the saturation limit. However, we confine our study to a regime where a disk exhibits continuous motion but with gradually decreasing speed.

When multiple camphor disks are present, the decrease in the activity of individual disks is proportionately faster. Besides, as the disks move closer, they interact via the surface tension fields of individual disks and tend to move away as the region between the disks is low in surface tension (or high in camphor concentration). In effect, this leads to a net repulsive interaction between the disks. Both the factors above are crucial in determining the dimensions of a dumbbell-shaped chamber so that the relevant dynamics are captured. In short, the chamber dimensions must be small enough to allow interactions between disks while simultaneously large enough for the disks to remain in continuous motion. At this point, we would like to highlight the three main experimental limitations of the current study: First, a considerable decrease in activity or aging of a disk over the experimental duration. Second, multi-disk systems age faster owing to the simultaneous release of camphor from multiple disks, reducing the experimental duration. Consequently, this results in poor dwell time statistics. Third, the dumbbell-chamber dimensions critically depend on the size and number of the disks. With these limitations, we studied the migration dynamics up to a four-disk system. In the following sections, we first discuss the results of single-disk dynamics followed by multi-disk dynamics.

\subsection{Single-disk dynamics}

\begin{figure}[h!]
\centering
\includegraphics[width=0.9\linewidth]{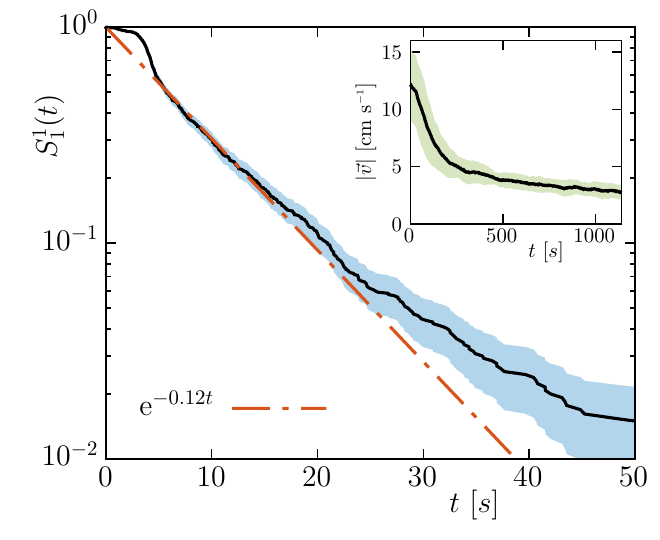}
\caption{The survival probability $S_{1}^{1}(t)$ for a disk to \emph{dwell} in a lobe showing a stretched-exponential decay with time. The black, solid line is the average $S_{1}^{1}(t)$ obtained over 13 trials, and the shaded area is the corresponding standard error. The inset figure shows the average speed $|\vec{v}|(t)$ (black, solid-line) of the camphor disk and the corresponding standard deviation (shaded area) over 13 trials. The decay in activity, i.e., $|\vec{v}|(t)$ in the current context, is reflected in the stretched-exponential behavior of survival probability.}
  \label{fig:figure3}
\end{figure}

We emphasize that a migration event of a disk from one lobe to the other is \emph{fundamentally} an escape event. In general, the escape of a particle through a narrow aperture of an otherwise closed and reflecting-wall-bounded region is known as the \emph{Narrow Escape Problem} (NEP) in literature \cite{Schuss2007-PNAS,Holcman2008-PRE,Debnath-JCP-2021}. An escape event is characterized by the total time taken for the escape process and is generally known as the \emph{First Passage Time} or the \emph{Escape Time}. Furthermore, it is established that the escape times (dwell times in the current context) are exponentially distributed in non-interacting particles, especially at large times, for various types \cite{Biswas-SM-2020, Kumar2023-SM, Olsen2020-PRR} of particles including chiral active particles which are a class of self-propelling entities with both translational and rotational degrees of freedom \cite{alakesh2024-SoftMatter}. As mentioned earlier, a camphor disk possesses both of these degrees of freedom and was recently modeled as chiral active particles \cite{Cruz2024-SoftMatter}. Since there is a nearly perfect symmetry between the two lobes of the dumbbell-shaped chamber, successive migrations can be regarded as multiple escape events from either lobe. Thus, we expect the survival probability also to decay exponentially. However, we observed that the survival probability decays as stretched-exponential. Figure~\ref{fig:figure3} shows the survival probability $S_{1}^{1}(t)$ (the black, solid line) and the corresponding standard error (the shaded area) for a single-disk system obtained over 13 trials. The dash-dotted (red-colored) line represents $S_{1}^{1}(t)$ fitted to an exponential. The observed survival probability progressively deviates from the exponential curve, suggesting a stretched exponential behavior. Intuitively, one can understand this behavior as follows. As we remarked earlier, the speed of a camphor disk gradually decreases with time, which results in longer dwell times at a later time in an experiment. This fact is chiefly reflected in the survival probability as well. The inset in figure~\ref{fig:figure3} shows the decrease in speed of a camphor disk with time. One may wonder whether survival probability and speed follow the same functional trend with time; but we observed otherwise. Following Cruz et. al.\cite{Cruz2024-SoftMatter}, we modeled a camphor disk as a chiral active particle and simulated the escape events from a circular domain with an aperture for a single-disk system. The simulation methodology and the results are discussed in~\ref{Appendix}. In brief, the survival probability always seems to be a stretched-exponential irrespective of the specific nature of activity (or speed) decay. Furthermore, we speculate that a decrease in activity with time (or \emph{aging}) is generally the cause of the stretched-exponential behavior in similar systems. For example, Bruckner et.al.~\cite{Bruckner2019-NP} investigated the migration of cancer cells in a rectangular-lobed dumbbell chamber and observed similar behavior. 

\subsection{Multi-disk dynamics}
\begin{figure}[h!]
\includegraphics[width=\linewidth]{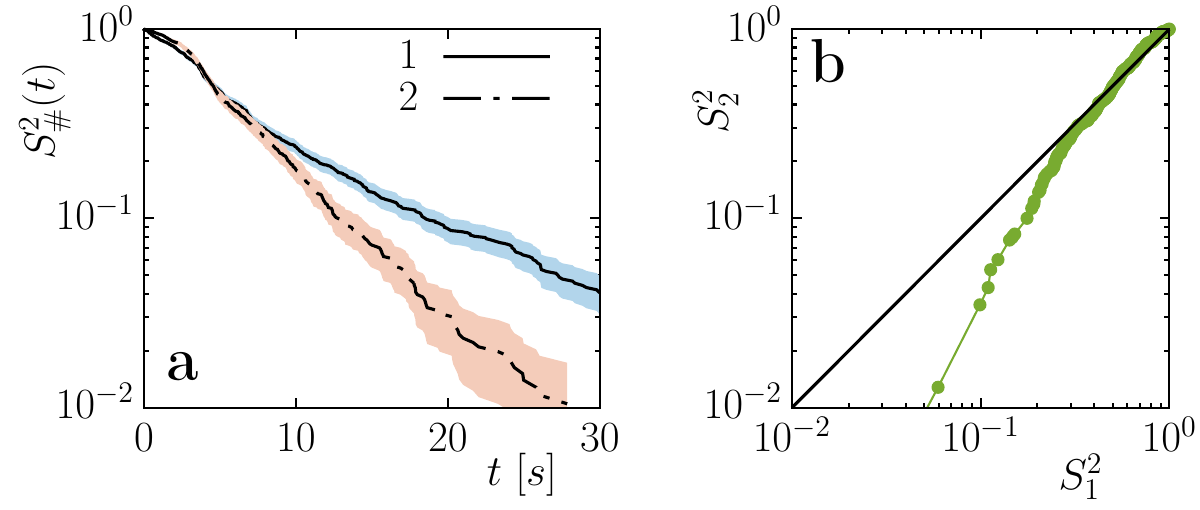}
\caption{(a) The survival probabilities $S_{1}^{2}(t)$, $S_{2}^{2}(t)$ for $\langle$\textbullet \textbar \textbullet$\rangle$, $\langle$\textbullet \textbullet \textbar$\rangle$ configurations respectively in a two-disk experiment. The lines represent average obtained over 13 trials and shaded areas represent the corresponding standard error. (b) The survival probabilities $S_{1}^{2}$ and $S_{2}^{2}$ plotted against each other. The solid, black line represents the line of equality.}
\label{fig:figure4}
\end{figure}

In this section, we investigated the impact of multiple disks on migration dynamics. Towards this, we conducted experiments with two (supplementary movie \emph{SM-2P.mp4}), three (supplementary movie \emph{SM-3P.mp4}), or four camphor disks (supplementary movie \emph{SM-4P.mp4}). The apparent effect in multi-disk systems is the decrease in speeds of individual disks as a result higher amounts of camphor adsorption on the water surface thereby reducing the surface tension gradients causing a decrease in thrust on the disks. Further, the migration dynamics of an individual disk remains unchanged qualitatively i.e., the survival probability remains a stretched-exponential with time. As mentioned earlier, the effect of interactions are best observed in survival probability of a configuration and is understood as follows. In a two-disk system, the repulsive interaction between the disks causes them to dwell apart such that the surface tension fields of the individual disk do not overlap. As a result, one would expect the $\langle$\textbullet \textbar \textbullet$\rangle$ to be a longer-living configuration than the $\langle$\textbullet \textbullet \textbar$\rangle$ and we observed the same. Figure~\ref{fig:figure4}a shows the survival probabilities $S_{1}^{2}(t)$ (solid, black line) and $S_{2}^{2}(t)$ (dash-dotted black line). Alternatively, we also plotted $S_{2}^{2}(t)$ \emph{vs.} $S_{1}^{2}(t)$ at the same time stamps in figure~\ref{fig:figure4}b which indicates that the $\langle$\textbullet \textbar \textbullet$\rangle$ configuration is more probable than the $\langle$\textbullet \textbullet \textbar$\rangle$ configuration at all times. Also, both the curves have a stretched exponential behavior for the same reasons explained in previous sections. 

\begin{figure}[h!]
\includegraphics[width=\linewidth]{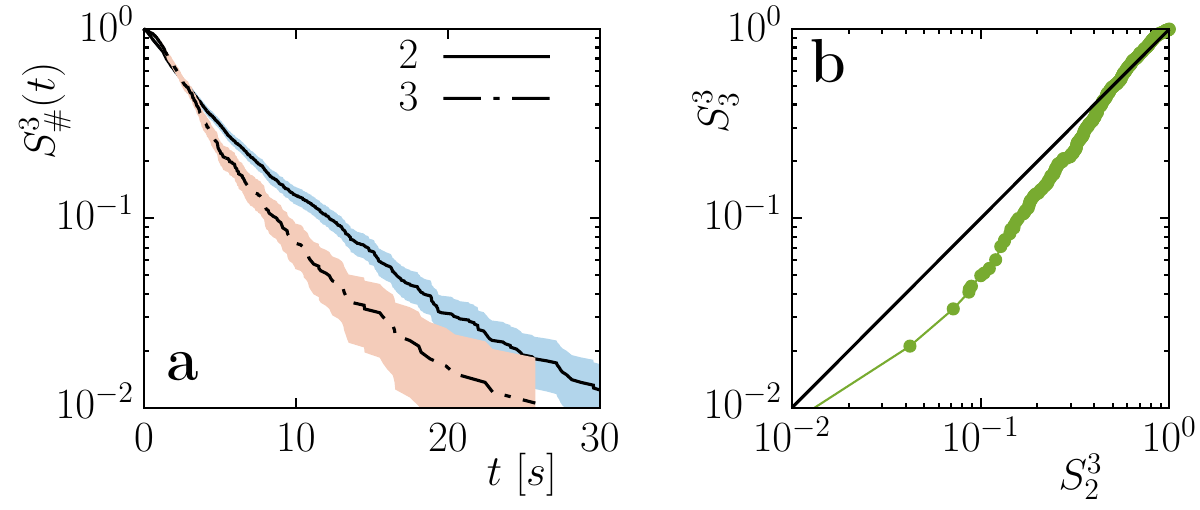}
\caption{(a) The survival probabilities $S_{2}^{3}(t)$, $S_{3}^{3}(t)$ for $\langle$\textbullet \textbullet \textbar \textbullet$\rangle$, $\langle$\textbullet \textbullet \textbullet \textbar$\rangle$ configurations respectively in a three-disk experiment. The lines represent average obtained over 13 trials and shaded areas represent the corresponding standard error. (b) The survival probabilities $S_{2}^{3}$ and $S_{3}^{3}$ plotted against each other. The solid, black line represents the line of equality.}
\label{fig:figure5}
\end{figure}

Now, for a three-disk system, the migration dynamics are qualitatively similar with $S_{2}^{3}(t)$ and $S_{3}^{3}(t)$ exhibiting stretched-exponential behavior (see figure~\ref{fig:figure5}a). Besides, the configuration $\langle$\textbullet \textbullet \textbar \textbullet$\rangle$ is more stable than the $\langle$\textbullet \textbullet \textbullet \textbar$\rangle$ configuration. Following the same rationale discussed earlier, two disks with one in each lobe experience minimal/negligible interaction. In contrast, the third disk occupying either lobe cannot avoid interacting with an existing disk, thus shuttles between the lobes. In other words, the third disk is ``\emph{frustrated}''. However, this configuration is more stable than $\langle$\textbullet \textbullet \textbullet \textbar$\rangle$, which has more number of pairwise interactions. In figure~\ref{fig:figure5}b, we plotted $S_{3}^{3}(t)$ against $S_{2}^{3}(t)$ and for most part the data falls below the line of equality at all times, once again suggesting $\langle$\textbullet \textbullet \textbar \textbullet$\rangle$ configuration is more stable.

\begin{figure*}[h!]
\includegraphics[width=1\linewidth]{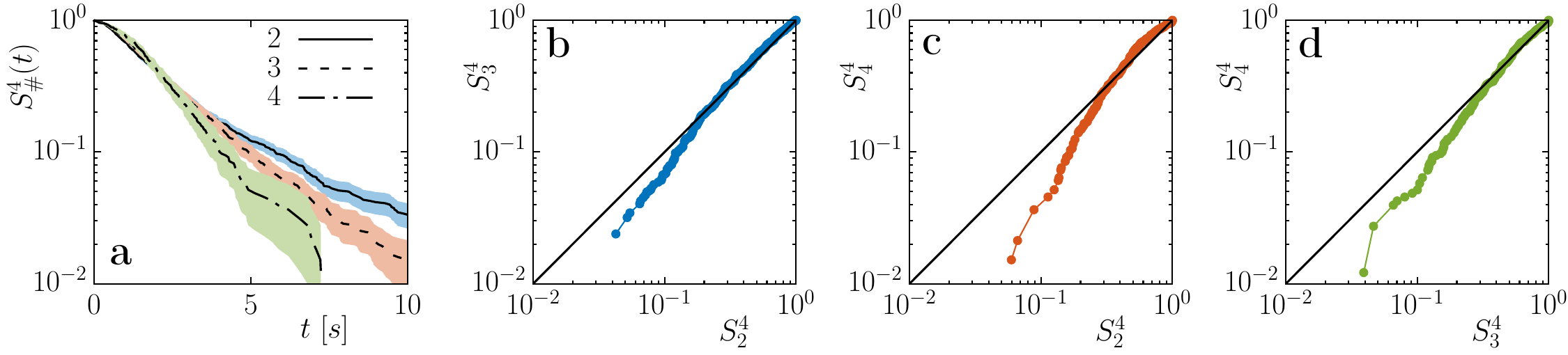}
\caption{(a) The survival probability, $S_{\#\mathrm{P}}(t)$ of two ($\#=2$), three ($\#=3$) or four ($\#=4$) disks to simultaneously \emph{dwell} in a chamber. The lines represent average obtained over 13 trials and shaded areas represent the corresponding standard error. The survival probabilities (b) $S_{\mathrm{2P}}$ \textit{vs.} $S_{\mathrm{3P}}$ (c)$S_{\mathrm{2P}}$ \textit{vs.} $S_{\mathrm{4P}}$, and (d) $S_{\mathrm{3P}}$ \textit{vs.} $S_{\mathrm{4P}}$.}
\label{fig:figure6}
\end{figure*}

Bearing the findings from two- and three-disk systems in mind, we make the following general statements limited to this study. The most stable configuration \emph{always} has the minimum number of pairwise interactions, followed by successive configurations with increasing pairwise interactions. Thus, in a $2N$-disk (even-numbered) system, the most stable configuration is the one with an equal number of disks in each lobe (i.e., $N$ in each). Similarly, in a $(2N+1)$-disk (odd-numbered) system, configuration with the disks equally distributed between the lobes but with one extra disk (i.e., $N$ in one lobe, $N+1$ in the other) shuttling between the lobes is the most stable configuration. In order to examine the above hypothesis, we studied the migration dynamics in a four-disk system. Once again, we calculated $S_{2}^{4}(t)$, $S_{3}^{4}(t)$ and $S_{4}^{4}(t)$ and the results are shown in figure~\ref{fig:figure5}. Once again, they all decay as stretched exponentials with time. Further, the order of decreasing stability is $\langle$\textbullet \textbullet \textbar \textbullet \textbullet$\rangle$ $>$ $\langle$\textbullet \textbullet \textbullet \textbar \textbullet$\rangle$ $>$ $\langle$\textbullet \textbullet \textbullet \textbullet \textbar$\rangle$ with the number of pairwise interactions for respective configuration following the opposite order, thus confirming the hypothesis.

\section{Conclusions}
\label{Conclusion}
In summary, we studied the migration dynamics of Marangoni surfers (camphor-infused paper disks) between two lobes of a dumbbell-shaped chamber. We characterized the dynamics using the survival probability of a given configuration wherein each lobe has a unique disk count. Qualitatively, all the survival probabilities exhibit stretched exponential decay with time\textemdash attributed to the disk's reducing activity/motility with time, which progressively increases the dwell times. We further modeled a camphor disk as a Chiral Active Particle and showed that the stretched exponential behavior is a direct consequence of reduced activity, and the exact functional nature of the activity decay is immaterial. Using the survival analysis of different configurations, we showed that the most stable configuration always has the least pairwise interactions, and the relative stability of the rest of the configurations decreases with increasing pairwise interactions. Generalizing this trend, we concluded that the most stable configuration always has an equal number of disks in each lobe: exactly half for an even-numbered system and one extra in either lobe for the odd-numbered system. Finally, we remark that exploring the migration dynamics of these surfers under heterogeneous conditions, such as different motilities, different lobe sizes, etc., is interesting and warrants further study.

\section*{Author Contributions}
AU and VSA conceived the experimental study; AU performed the experiments and analyses; VSA prepared the figures; AU prepared the original draft; VSA reviewed and edited the final draft.

\section*{Conflicts of interest}
There are no conflicts to declare.

\section*{Acknowledgements}
The research work is supported by Ministry of Human Resources Development of India. Computing time and resources provided by the High Performance Computing facility Agastya at the Indian Insitute of Technology Jammu is gratefully acknowledged.

\appendix
\renewcommand{\thesection}{Appendix \Alph{section}}
\renewcommand{\thefigure}{\Alph{section}\arabic{figure}}
\setcounter{figure}{0}
\section{Camphor Boat as a Chiral Active Particle}
\label{Appendix}

\begin{figure}[h]
\includegraphics[width=0.9\linewidth]{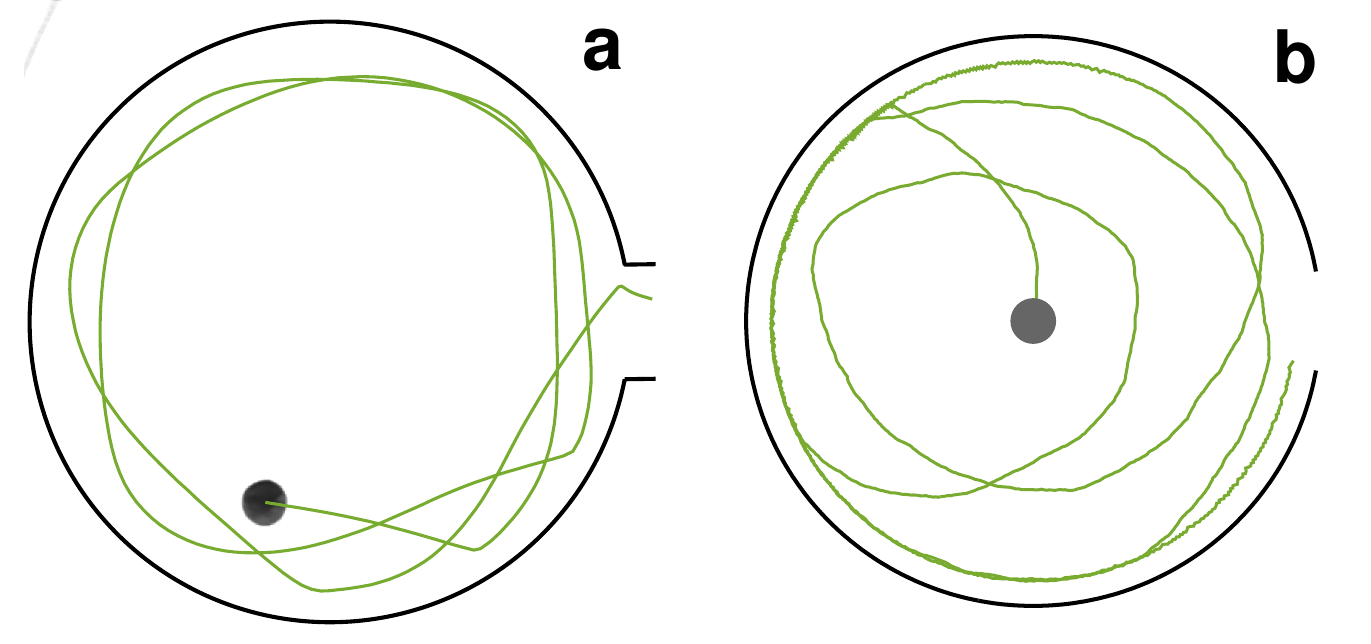}
\caption{(a) Trajectory of a camphor-infused paper disk from experiments (contrast adjusted and dumbbell boundary highlighted for clarity). (b) Trajectory of a Chiral Active Particle (CAP) from simulations. }
\label{fig:figure7}
\end{figure}

Following Cruz et al. \cite{Cruz2024-SoftMatter}, we model a Marangoni surfer (the camphor-infused paper disk) as a Chiral Active Particle (CAP) to study the effect of aging or activity (speed of the disk) decay on survival probability. By exploiting the symmetry between the two lobes of the dumbbell-shaped chamber (established in section~\ref{Methods}), we \emph{only} simulate the escape events from a single lobe, i.e., a circular domain with an aperture for a single disk system. We emphasize that the model is minimalistic and captures \emph{only} the qualitative behavior of the surfer's motion. Moreover, it only involves a partial physical description of the complex underlying processes that result in the surfer's motion. In brief, the surfer's motion is governed by the overdamped Langevin's equations:

\begin{subequations} \label{eq:eom}
\begin{align} 
\dot{\vec{r}} &= v \hat{\theta} + \sqrt{2\mathrm{D}_{T}}\vec{\xi}(t) \label{eq:eomt} \\
\dot{\theta} &= \omega + \sqrt{2\mathrm{D}_{R}}\zeta(t) \label{eq:eomr}
\end{align}  
\end{subequations}

Where $\vec{r}(t)$ and $\theta(t)$ are the instantaneous position and orientation of the particle on the water surface, in addition, $\hat{\theta} = (\cos{\theta(t)}, \sin{\theta(t)})$ is the unit vector in the direction of instantaneous orientation. $\vec{\xi} = (\xi_{x}, \xi_{y})$ and $\zeta$ are Gaussian white noise variables with zero mean and unit variance. These noise levels are arbitrarily chosen such that the simulated trajectories qualitatively agree with the experimentally observed trajectories (see Figure~\ref{fig:figure7}). Further, the radius of the domain $R = 35\ \mathrm{\mu m}$, the translational speed $v = 31.4\ \mathrm{\mu m\ s^{-1}}$ and rotational speed $\omega = 1.346\ \mathrm{rad\ s^{-1}}$ are fixed and are chosen such that the non-dimensional combination of the three quantities i.e., $\omega R/v \sim 1.5$ is in \emph{scaled congruence} with the experiments. It is worth noting that the quantity $\omega R/v$ is the ratio of the domain radius $R$ and the radius of trajectory $v/\omega$ of the CAP. One more detail worth mentioning is that the camphor disk's diameter is $6\ \mathrm{mm}$, and hence, the closest a disk can approach the dumbbell boundary is $\sim 3\ \mathrm{mm}$ due to the excluded volume effects. In contrast, the CAP is a point particle and can wander arbitrarily close to the domain boundary. Therefore, the CAP's trajectories are drawn (figure~\ref{fig:figure7}b) by enlarging the domain size by $3\ \mathrm{mm}$ to visualize the similarity of trajectories. Finally, we employ Euler-Maruyama scheme with reflective boundary conditions at the dumbbell boundary to numerically integrate equations~\eqref{eq:eom} and obtain the dwell time statistics and the corresponding survival function. 

\begin{figure}[h]
\includegraphics[width=1\linewidth]{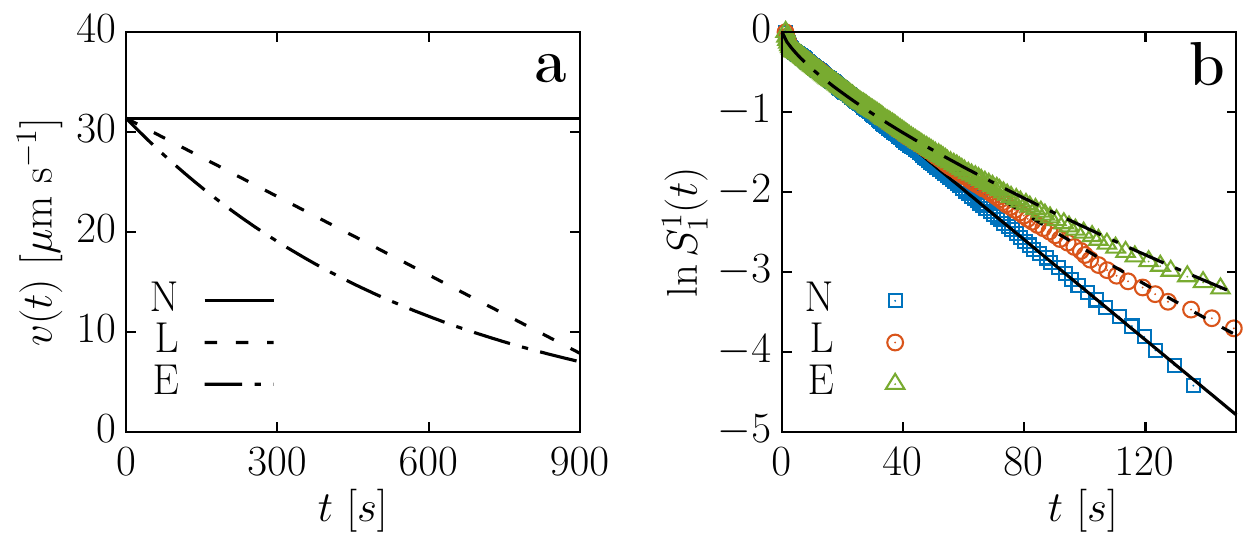}
\caption{(a) The speed $|\vec{v}|(t)$ of a Chiral Active Particle (CAP) versus time in three different cases namely 1. No decay (N) 2. Linear decay (L) and (3) Exponential decay (E); (b) the survival functions $S_{1}^{1}(t)$ for the corresponding cases. The \textit{No decay} case exhibits exponential tail in survival function whereas the other two cases exhibit stretched exponential behavior; as evident from the fitted curves.}
\label{fig:figure8}
\end{figure}

Using the survival analysis in three different scenarios \textit{namely} no decay (N), linear decay (L) and exponential decay (E) in speed, we demonstrate that the decay in activity (or decrease in speed) of the particle causes the stretched exponential behavior. Further, the stretched exponential nature of survival function is independent of the nature of activity decay or aging. Figure~\ref{fig:figure8}a shows the speed $|\vec{v}|(t)$ of the particle with time, $t$ in the three different scenarios and the corresponding survival functions $S_{1}^{1}(t)$ are shown in figure~\ref{fig:figure8}b. Clearly, the survival function is exponentially tailed in the no decay (N) case~\cite{alakesh2024-SoftMatter} and is a stretched exponential in the linear (L) and exponential (E) cases. 

%%%END OF MAIN TEXT%%%

%The \balance command can be used to balance the columns on the final page if desired. It should be placed anywhere within the first column of the last page.

\balance

%If notes are included in your references you can change the title from 'References' to 'Notes and references' using the following command:
%\renewcommand\refname{Notes and references}

%%%REFERENCES%%%
\bibliography{manuscript} %You need to replace "manuscript" on this line with the name of your .bib file
\bibliographystyle{manuscript} %the manuscript's .bst file

\end{document}